%% file: main.tex
\documentclass{article}
\input{packages}
\input{macros}
\title{Time transport correlations in abelian sandpile models}
\author{Valentin Lallemant$^1$\\
$^1$\small{\textit{Université Grenoble Alpes, CNRS, LPMMC, 38000 Grenoble, France}}}
\date{December 2025}

\begin{document}

\maketitle
\begin{abstract}
	Sandpiles form one of the largest class of models displaying a critical stationary state.
	Despite a few decades of research, a comprehensive and systematic rigorous characterisation of their spatial and, even more, time dependent properties has remained elusive. 
	Among the obstacles, we can mention their out of equilibrium and non-linear dynamics features which prevent, in general, the access to the stationary properties explicitly.
	In fact, even the knowledge of the stationary state is quite exceptional in sandpiles.
	In that respect, it has become standard to develop a model to model strategy and, so to say, general results or tools applicable to these systems are missing.
    In this paper, we unveil general and simple properties of time transport correlations in certain classes of abelian sandpile models.
    We proceed gradually, starting from results applicable in a broad context, to more and more specific ones, consequently valid to smaller and smaller classes.
    For instance, we show, under a few hypothesis, that the number of particles dissipated displays mostly anticorrelation in time. 
    Besides, on a more integrable point of view, the approach followed might culminate with the proof of a link between $2$-points time transport correlations and the second moment of the integrated transport over time.
    To be clear, these two quantities are related through a linear system of equations which is explicitly solved and applies to at least three $1$D sandpile models, namely the Directed Stochastic Sandpile, the Oslo and the Activated Random Walk (in a peculiar setup) models. 
\end{abstract}

\input{introduction}
\input{definitions}
\input{centralresult}
\input{boundedvar}
\input{application}
\input{conclusion}
\printbibliography[title={Bibliography}]
\end{document}

%% file: packages.tex
\usepackage{array} %Required for making cell with diagonal separation in tabular
\usepackage{amsthm}
\usepackage{amsmath}
\usepackage{amsfonts}
\usepackage{amssymb}
\usepackage{graphicx} %Required for inserting images
\usepackage{mathtools}
\usepackage{stmaryrd}
\usepackage{xcolor}
\usepackage{dsfont}
\usepackage{hyphenat}
\usepackage{hyperref}
\usepackage{forest}
\usepackage{tikz}
\usetikzlibrary{arrows,patterns}
\usetikzlibrary {arrows.meta,bending,positioning}
\usepackage{accents}
\usepackage{titlesec}
\usepackage{caption} %Allow to specify caption location 
\usepackage{nicematrix} % Required for making nice tabular

\usepackage[backend=bibtex, style=alphabetic,]{biblatex} % Use the bibtex backend with the authoryear citation style (which resembles APA)

%% file: macros.tex
\newcommand\avrg[1]{\langle#1\rangle}
\newcommand\bra[1][]{\langle#1|}
\newcommand\configurations{\mathcal{G}}

\newcommand\evolution{\mathcal{E}}
\newcommand\evol{E}
\newcommand\Id{\mathbb{I}} %Identity
\newcommand\intrange[1][]{\llbracket #1\rrbracket} %Integer interval
\newcommand\ket[1][]{|#1\rangle} %ket notation

\newcommand\recur{\mathcal{R}}
\newcommand\stable{\mathcal{C}}
\newcommand\setpaths{\mathcal{A}}
\newcommand\stat{\mathcal{S}} %Vector space of the statistical state
\newcommand\zcfg[1]{|#1\rangle}
\newcommand\Z[1][]{\mathbb{Z}^{#1}}

\newtheorem{lemma}{Lemma}
\newtheorem{proposition}{Proposition}
\newtheorem{remark}{Remark}

\newtheorem{definition}{Definition}

%% file: introduction.tex
\section{Introduction}

This work comes within the scope of the general concept of Self-Organised Criticality (SOC).
Originally, and as the very name of the father paper \cite{bak_self-organized_1987} of SOC suggests, the concept provides a generic mechanism under which power law behavior, i.e. long range correlations, emerges in observables of out of equilibrium systems. 
In that respect, the stationary state of these systems can be qualified to be critical, in analogy with the phenomenology that develops in equilibrium systems at a (continuous) phase transition point. 
The mechanism proposed in \cite{bak_self-organized_1987} can be viewed simply as the capacity of a system extended in space to accumulate, redistribute (transport) and dissipate local stresses under some conditions. 
In fact, the accumulation is usually feasible up to a certain threshold value which, when exceeded, triggers either transport and/or dissipation of the stress.
Dissipation can, for instance, happen when the stress reaches a specific zone in space.
It is now consensually admitted that many instances, i.e. models, falling within these simple statements display what was initially intended to be, meaning a critical stationary state.
In particular, one of the main power law behaved observables are associated to transport in the systems.

The comprehension of sandpile models have benefited from a few decades of research with a tremendous range of techniques and approaches. 
We cannot report here a complete description of this literature (\cite{Pruessner_2012} would be a good starting point), but we can comment on some of the achievements and issues that have been, and still, are encountered. 
First of all, the development of the computational power resources have permit to investigate most of the sandpiles using simulations (e.g. \cite{grassberger_oslo_2016}) and exact numerical methods (e.g. \cite{corral_calculation_2004}). 
One other approach has been to try understanding the systems analytically through various approximation schemes and hypothesis, of which the field theoretical formulation might be the most prominent one (e.g. see the review \cite{wiese_theory_2022}). 
Finally, the last point of view adopted is the one of rigorous results.
This is probably the less successful approach in terms of power of prediction, of which the lack of general tools is one fundamental problem.
After stating the obvious, we emphasize that the present work is an attempt toward developing general rigorous strategies and/or to propose another intuition on certain questions around sandpile models.

More particularly, the problem that we address here is about time correlations of transport observables in abelian sandpile models.
This question have only but a few works on its account, most of which have been using numerical (simulations) or field theoretic arguments (see \cite{pradhan_time-dependent_2021}).
Among the rigorous results, we can cite at least \cite{dhar_steady_2004, garcia-millan_correlations_2018} which are of a particular importance for the results of the present paper.
We must underline that this very topic of time correlations in sandpile has been regularly regarded as one interesting open direction of research (\cite{jonsson_area_1998, dhar_theoretical_2006, levine_universality_2023}).
One obstacle identified around this question has to do with the interplay between the dynamical process, the avalanches, and the content of the system, the particles, which, by construction of sandpile models, influence each other.
For instance, the space correlations that might develop in sandpile stationary states are to be conceived precisely as the consequence of this interaction avalanche/density of particles. 
On the other hand, describing an avalanche along its propagation is a well-known difficult task, even for small system sizes and $1$D sandpiles.
These two difficulties combined together prevent, in nearly all cases, to express the stationary state in an explicit form. 
Since the stationary state is a prerequisite for computing stationary observables (static or transport ones), it seems therefore hopeless to derive rigorous results on the system observables without it (apart from some very peculiar observables benefiting global constraints).

As have been already said, in the present paper, we provide answers to a variety of problems around time transport correlations in sandpile models. 
We try, along the discussion, to go from the most general results, applicable to a large class of sandpiles, and gradually increase the hypothesis, at the same time narrowing the study to smaller and smaller classes. 
The first section introduces a general definition of what we mean by abelian sandpile models.
At the end of it, we introduce $4$ different models known from the literature and upon which specific results of the paper will be shown to be applicable.
In the second section we exhibit a central recursion relation inherited from the abelian nature of the sandpile under consideration.
In the third one, we discuss of some generic constraints satisfied by some transport observables. 
This part is seen as an extension of the work of \cite{garcia-millan_correlations_2018}. 
In the fourth section, we apply the previous results to a specific class of sandpiles and manage to connect via a linear system of equations two different current properties, namely time transport correlations and the second moment of the integrated transport over time. 
Lastly, we conclude the article with a sum up of the results and comment on future direction of research.

%% file: definitions.tex
\section{Definitions}\label{section: definitions}

We start giving a general definition of the sandpile models considered in this paper.

Take a graph $G(V,E)$. 
At each vertex $v\in V$ are attached two positive discrete fields $(w(v),z(v))$. 
$w(v)$ counts the number of local waiting particles, whereas $z(v)$ counts the number of stable particles. 
A waiting particle evolves in time whereas stable particles do not.
A configuration $c$ at a given time of the evolution is fully represented by the couple $c\sim(w,z)$.
Supposing the graph $G(V,E)$ fixed, we denote by $\configurations$ the set of all finite configurations, i.e. all $c\sim(w,z)$ such that $\forall v\in V$, $w(v),z(v)<+\infty$.
We also denote by $\stable$ the set of stable configurations, i.e. the subset of $\configurations$ for which any $c\sim(w,z)$ has the additional property $\forall v\in V$,  $w(v)=0$, meaning $c$ does not contain waiting particles.
We will restrain ourselves to the case where $\stable$ is finite and scales with the number of nodes $|V|$. 
To do so, we forbid any stable configuration to have one vertex $v\in V$ where $z(v)>z_{\max}(v)$ where $z_{\max}(v)$ is a threshold which can be vertex dependent or constant over the system.
In addition, we will mostly use the configuration representation 
\begin{equation}\label{eq: notation generalised configuration}
	(w,z)=\bigotimes_{v\in V}\big(a_v^{w(v)}\ket[z(v)]\big)
\end{equation} 
which separates at a glance the waiting part from the stable one. 
By construction $a_v$ represents one waiting particle at site $v\in V$.
In this context, a state is a linear combination of configurations with positive weights whose sum equals to $1$ (normalised statistical state). 
We call by $\stat:=\textrm{span}(\configurations)$ the corresponding vector space.

In the sandpiles considered, the dynamical process between the microscopic time $t$ and $t+1$ is described by a Markov operator $\evol:\stat \to \stat$.
The action of this operator is defined for each vector $c\in\configurations$, constituting the $\stat$ basis, by
\begin{equation}\label{eq: action evol}
	E(c)= \sum_{c'\in\configurations}P(c\to c')c'
\end{equation}
where $\sum_{c'\in\configurations}P(c\to c')=1$. 
We suppose also that $\evol$ acts locally.
This assumption is not a necessary condition for most of the results of the paper to be true.
Nonetheless, this feature is extremely common in the literature of sandpiles and we prefer to include its description in the section.
Locality here is understood as the fact that \emph{only one waiting particle}, and afterward the local stable part around, are evolved by its action.
Among the new configurations $c'$ of Equation \eqref{eq: action evol}, we can have (locally) lost, moved particles and/or activated stable particles . 
We say that a site/ vertex/ node have performed one toppling when $\evol$ have moved at least one particle from the vertex to another one.
The condition that triggers the different outcomes depends only on a the local value of the stable field $z$.
For example, one can consider the gradient of stable particles in one direction ($\sim$ grains in the Oslo model), or their total number at a specific vertex ($\sim$ Activated Random Walk model (ARW)).
Also, since $\evol$ is conditioned on evolving a waiting node, we have necessarily 
\begin{equation}
	\evol(c)= c
\end{equation}
for all $c\in\stable$.
In fact, sandpile models were originally conceived as models undergoing an absorbing phase transition.
The absorbing (=stable) phase is guarantied as long as, for all $c\in\configurations$, 
\begin{equation}\label{eq: absorbing phase condition}
	\evol^{\infty}(c)=\sum_{c'\in\stable}P^{\infty}(c\to c')c'
\end{equation}
where $\evol^{k}(c)=\evol \circ...\circ \evol(c)$ is the iterate action of $\evol$ on $c$ $k$ times.
Equation \eqref{eq: absorbing phase condition} corresponds to \emph{the stabilisation} of the initial state $c$.

On top of what have been discussed, we restrain the discussion on sandpiles with \emph{abelian stabilisation}. 
The abelianess is understood as following: instead of a unique operator, consider a whole set of operators $\evolution$.
Two operators $\evol,\evol'\in\evolution$ are different if there exists $c\in\configurations$ for which $\evol(c)\neq\evol'(c)$.
The difference between $\evol$ and $\evol'$ is attributed to a difference in the order in which waiting particles are evolved.
The abelianess is then conceived as this difference in the order being irrelevant for the stabilised distribution generated. 
In equation this reads
\begin{equation}\label{eq: abelian relation}
	\evol^{\infty}(c)=\evol'^{\infty}(c)
\end{equation}
for any $\evol,\evol'\in\evolution$.

In the next sections, we will be interested in the transport properties when starting from a stationary distribution.
Stationarity is ensured as long as particles are injected and dissipation (of particles) occurs in the system.
More, for the sandpile to have an intermittent dynamic with well defined avalanches, we trigger the injection of a particle only after the previous state has stabilized, and stabilize the new excited state.
Adding a particle and stabilizing is considered as a (macroscopic) time step $t \to t+1$. 
To simplify the discussion, we will suppose the injection to occur on a unique site that we label $1$ or $v_1\in V$.
To be more precise, take $c\sim(w,z)\in \configurations$.
Using the notations \eqref{eq: notation generalised configuration}, during the next time step, we stabilize the configuration $a_1c$ where $a_1c\sim(w+\delta_{v,1},z)$. 
Iterating this scheme, we are therefore interested in the Markov process 
\begin{equation}
	W:=\evol^{\infty}\circ a_1
\end{equation}
where the separation of time scales between the stabilisation and the injection is ensured.
In particular, the stationary state, which we denote $\ket[\psi]$, satisfies 
\begin{equation}
	W\ket[\psi] = \ket[\psi]=\sum_{r\in\recur}P(r)\ket[r]
\end{equation}
where $\recur \subset \stable$ is the set of recurrent configurations, by which we mean stable configurations which appear with non zero probability in the stationary state.
From this, it is obvious that for any $r\in\recur$, we have $W\ket[r]=\sum_{r'\in\recur}P(a_1 r \to r')\ket[r']$.

Anticipating the lengthy expressions appearing when dealing with transport observables, we now specify a number short notations.
First, we denote by $u(\tau):=\sum_{t=1}^{\tau}u_t$ the super response triggered by $\tau$ successive particles $a_1$ added in the system.
$u_t$ corresponds, in a given sequence of avalanches, to the value of the $u$ observable of the $t$-th avalanche triggered.
$u(\tau)$ is a priori a sum of correlated random variables.
If $\tau =1$, we simplify the notation $u(1)=u$.
In addition, we will first suppose that the transport observable $u$ along an avalanche takes a unique value $u(c\to c')$ that depends only on the initial and final configurations $c,c'\in \configurations$.
We discuss afterward the case where the value of $u$ is not unique.
In that context, we use the following shortcuts
\begin{align*}
\avrg{.}&:= \bra[\Id].\ket[\psi]\\
\avrg{u}=\bra[\Id]u\ket[\psi]&=\sum_{r,r'\in\recur}P(r)P(a_1r\to r')u(a_1r\to r')\\
\avrg{W^t}&=\sum_{r,r'\in\recur}P(r)P(a_1^t r\to r')\\
\avrg{u^k(t_3)W^{t_2}u(t_1)}&=\sum_{r,r',r'',r'''\in\recur}P(r)P(a_1^{t_1}r\to r')\times \\
&u(a_1^{t_1}r\to r')P(a_1^{t_2}r'\to r'')P(a_1^{t_3}r''\to r''')(u(a_1^{t_3}r''\to r'''))^k
\end{align*} 
where $u$ can be any transport observable.
One can see that inside of the average symbol $\avrg{.}$ we use the convention of ordering in causal order, or increasing time, from right to left the observables.
One should be careful that the notation $u^k(\tau)$ and $W^k$ inside of $\avrg{.}$ have different meaning, respectively to evaluate the $k$-th moment of $u(\tau)$ and to evolve $k$ times an initial state performing $k$ successive avalanches.
When we need to evaluate correlation in time for $u$, we resort to the notation $\avrg{uW^tu}$ and, for $t=0$, set $\avrg{uW^0u}=\avrg{uu}\not = \avrg{u^2}$.
In the case where $u$ is not unique for a given transition $c\to c'$ with $c,c'\in \configurations$, we denote by $\setpaths(c\to c')$ the set of all paths/ avalanches from $c$ to $c'$. 
We then have to replace $P(c\to c')u(c\to c')$ by $\sum_{v\in\setpaths(c\to c')} P(v)u(v)$ where $P(c\to c'):= \sum_{v\in\setpaths(c\to c')} P(v)$.
 
We end this section introducing $4$ abelian models selected from the literature on sandpiles. 
In fact, the models have been chosen for their irreversible nature. 
A more detailed discussion will be provided along the argumentation about this point. 
\begin{definition}[Directed Stochastic Sandpile (DSS) \cite{pruessner_exact_2004}]\label{def: DSS}
	The DSS model in $1$D describes the evolution of particles over the linear chain $V=\intrange[1,L]$ made of $L$ sites.
	The evolution is given by	
	\begin{align*}
		&\evol(a_x\zcfg{1})= a_x a_{x+1}\zcfg{0}\\
    		&\evol(a_x\zcfg{0})= p\zcfg{1}+qa_{x+1}\zcfg{0}
	\end{align*}
	for some $\evol\in\evolution$ and $x\in\intrange[1,L]$. 
	Injection occurs at $x=1$ and a right dissipative boundary is set so that $W=\evol^\infty\circ a_1$ and $a_{L+1}c=c$ for all $c\in\stable$.
\end{definition}
\begin{definition}[Oslo \cite{christensen_tracer_1996}]\label{def: Oslo}
	The Oslo model in $1$D describes the evolution of particles over the linear chain $V=\intrange[1,L]$ made of $L$ sites.
	The evolution is given by	
	\begin{align*}
    		&\evol(a_x\zcfg{0})= \zcfg{1}\\
    		&\evol(a_x\zcfg{1})= p\zcfg{2}+qa_{x-1} a_{x+1}\zcfg{0}\\
    		&\evol(a_x\zcfg{2})= a_{x-1} a_{x+1}\zcfg{1}\\
	\end{align*}
	for some $\evol\in\evolution$ and $x\in\intrange[1,L]$. 
	Injection occurs at $x=1$ and a left dissipative and right reflexive boundaries are set so that $W=\evol^\infty\circ a_1$, $a_0c=c$ and $a_{L+1}c=a_Lc$ for all $c\in\stable$.
\end{definition}
\begin{definition}[Activated Random Walk (ARW)]\label{def: ARW}
	The ARW in $1$D describes the evolution of particles over the linear chain $V=\intrange[1,L]$ made of $L$ sites.
	The evolution is given by	
	\begin{align*}
    		&\evol(a_x\zcfg{0})= p\zcfg{1}+\frac{q}{2}(a_{x-1}\zcfg{0}+a_{x+1}\zcfg{0})\\
    		&\evol(a_x\zcfg{1})= \frac{1}{2}(a_{x-1}a_x\zcfg{0}+a_xa_{x+1}\zcfg{0})
	\end{align*}
	for some $\evol\in\evolution$ and $x\in\intrange[1,L]$. 
	The driven-dissipative setup that we choose here consists in injecting particles at $x=1$ and to have a left reflexive and right dissipative boundaries so that $W=\evol^\infty\circ a_1$, $a_{L+1}c=c$ and $a_0c=a_1c$ for all $c\in\stable$.
\end{definition}
This history of the ARW is a bit convoluted. 
We refer to \cite{rolla_absorbing-state_2012} for some comments on its genesis.

\begin{definition}[Manna (with center of mass conservation) \cite{Manna_1991}]\label{def: Manna}
	The Manna  model in $d\geq 2$ dimension describes the evolution of particles over the subdomain of $V\subset \Z[d]$ such that $|v_1-v|=L-1$ is the distance between the injection point $v_1$ and all the sites on the boundary $v\in \overline{\partial V}\subset V$.
	The distance here is understood as the minimal number of edges separating two vertices.
	In that sens, $v_1$ is the central site of the system.
	The evolution is given by	
	\begin{align*}
    		&\evol(a_v\zcfg{0})= \zcfg{1}\\
    		&\evol(a_v\zcfg{1})= \frac{1}{d}\sum_{i=1}^{d}a_{v-\delta_i}a_{v+\delta_i}\zcfg{0}
	\end{align*}
	for some $\evol\in\evolution$ and $v\in V$. 
	Besides, $\delta_i$ is the positive unit vector on $\Z[d]$ in the $1\leq i\leq d$ direction. 
	Injection is performed on the central site $v_1$ and dissipation occurs when particles leave $V$.
	Equivalently, particles are dissipated when they reach any $\tilde{v}\notin V$ for which $\exists v\in V$ such that $|v-\tilde{v}|=1$.
\end{definition}

%% file: centralresult.tex
\section{Preliminary results}

\begin{proposition}[Abelianess and linearity of integrated transport properties]\label{prop: abelian is linearity}
Take any abelian sandpile. 
Then, starting from the stationary state, and for any operator $u(\tau)=\sum_{t=1}^{\tau}u_t$ quantifying a transport property (local or global) accumulated over $\tau$ avalanches, we have
\begin{align}\label{eq: lineartiy of <u(tau)>}
    \avrg{u(\tau)} &= \sum_{t=1}^{\tau}\avrg{W^{\tau-t}uW^{t-1}}=\sum_{t=1}^{\tau}\avrg{u}
\end{align}
with $u(1):=u$
\end{proposition}
\begin{proof}
    This is rather straightforward and we treat the general case where the transport quantity is not unique, in general, for a given transition $a_1 r\to r'$ with $r,r'\in\recur$ (avalanches in ARW for example).
    Denote by $\setpaths(a_1r\to r')$ the set of all the different avalanches/paths that perform the transition $a_1r$ to $r'$.
    By definition we have 
    \begin{align*}
        \avrg{u(\tau)} &= \sum_{r,r'\in\recur}P(r)\sum_{v\in \setpaths(a_1^{\tau}r\to r')} P(v)u(v)\\
        &=\sum_{r\in\recur}P(r)\sum_{v\in \setpaths(a_1^{\tau-1}r\to r')}P(v)\sum_{v'\in \setpaths(a_1r'\to r'')}P(v')(u(v)+u(v'))\\
        &= \sum_{r\in\recur}P(r)\sum_{v\in \setpaths(a_1^{\tau-1}r\to r')}P(v)u(v)\sum_{v'\in \setpaths(a_1r'\to r'')}P(v')\\
        &+\sum_{r\in\recur}P(r)\sum_{v\in \setpaths(a_1^{\tau-1}r\to r')}P(v)\sum_{v'\in \setpaths(a_1r'\to r'')}P(v')u(v')\\
        &=\avrg{Wu(\tau-1)}+\avrg{uW^{\tau-1}}
    \end{align*}
    Iterating the recursion on the second term in the r.h.s. and remarking that $\avrg{u W^{t}}=\avrg{u}$ for all $t$, as we start from the stationary state, yield the final result \eqref{eq: lineartiy of <u(tau)>}. 
\end{proof}
The average response of an initial super avalanche triggered by several particles is then linear in the number of particles added.
In fact, the first equality of \eqref{eq: lineartiy of <u(tau)>} is more fundamental as it is true for any initial state, i.e. is valid when replacing $\avrg{.}=\bra[\Id].\ket[\psi]$ with $\avrg{.}=\bra[\Id].\ket[\phi]$ for any $\phi\in\stat$.
This simple property is an important constraint for the system time correlations as we will see.

\begin{lemma}[General recursive relation for time transport correlations]\label{lem: recursion transport corr of ASM}
   	Take any abelian sandpile model in its stationary state and any transport observable $u$.
    Then we have the recursive relation for any $\tau_1,\tau_2\in N$  
    \begin{align}\label{eq: recursion transport corr of ASM}
    		\avrg{u^2(\tau_1+\tau_2)}-\avrg{u(\tau_1+\tau_2)}^2 
	=\avrg{u^2(\tau_1)}-\avrg{u(\tau_1)}^2 &+\avrg{u^2(\tau_2)}-\avrg{u(\tau_2)}^2+ 2f(\tau_1,\tau_2)
	\end{align}
	with
	\begin{multline}
	f(\tau_1, \tau_2) = \sum_{k=0}^{\theta-1}(k+1)\avrg{uW^{k}u}
	+ \theta\sum_{k=\theta}^{\theta'-2}\avrg{uW^{k}u}\\
		+\sum_{k=\theta'-1}^{\tau_1+\tau_2-2}(\tau_1+\tau_2-1-k)\avrg{uW^{k}u}
	-\tau_1 \tau_2\mu^2
    \end{multline}
    where $\mu:=\avrg{u}$, $\theta = \min(\tau_1,\tau_2)$ and $\theta' = \max(\tau_1,\tau_2)$.
\end{lemma}
\begin{proof}
	For simplicity, we suppose first that any transition $a_1r\to r'$ is unique, and cover the more general situation when there is degeneracy at the end.
	The second term in the l.h.s. of \eqref{eq: recursion transport corr of ASM} gives
	\begin{equation*}
		\avrg{u(\tau_1+\tau_2)}^2 = (\sum_{t=1}^{\tau_1+\tau_2}\mu)^2 = (\tau_1+\tau_2)^2\mu^2
	\end{equation*}
	Instead, for the first term, we have
	\begin{align*}
	\avrg{u^2(\tau_1+\tau_2)}  &= \sum_{r,r'\in\recur}P(r)P(a_1^{\tau_1+\tau_2} r\to r')(u(a_1^{\tau_1+\tau_2} r\to r'))^2\\
	&= \sum_{r,r',r''\in\recur}P(r)P(a_1^{\tau_1} r\to r')P(a_1^{\tau_2}r'\to r'')(u(a_1^{\tau_1} r\to r')+u(a_1^{\tau_2}r'\to r''))^2 \\
	&=\avrg{u^2(\tau_1)}+\avrg{u^2(\tau_2)}+2\avrg{u(\tau_2)u(\tau_1)}
	\end{align*}
	One can also rewrite 
	\begin{align*}
	\avrg{u(\tau_2)u(\tau_1)} &=\avrg{\sum_{t=0}^{\tau_2-1}W^{\tau_2-t-1}uW^{t}\sum_{t'=0}^{\tau_1-1}W^{t'}uW^{\tau_1-t'-1}}\\
	&=\sum_{t=0}^{\tau_1-1}\sum_{t'=0}^{\tau_2-1}\avrg{uW^{t+t'}u}\\
	&=\sum_{k=0}^{\theta-1}(k+1)\avrg{uW^{k}u}
	+ \theta\sum_{k=\theta}^{\theta'-2}\avrg{uW^{k}u} +\sum_{k=\theta'-1}^{\tau_1+\tau_2-2}(\tau_1+\tau_2-1-k)\avrg{uW^{k}u}
	\end{align*}
	with $\theta = \min(\tau_1,\tau_2)$ and $\theta' = \max(\tau_1,\tau_2)$.
	Since $W$ is a stochastic matrix and we start from the stationary state, we again used the relation $\avrg{W^ku}=\avrg{u}=\avrg{uW^{k'}}$ satisfied for any $k,k'\geq 0$.
	
	We now go back to the case where all the avalanches $a_1r\to r'$ do not share a unique value of $u$.
	In that case, recalling that $\setpaths(a_1r \to r')$ is the set of paths associated to the transition $a_1r \to r'$, we have 
	\begin{align*}
		\avrg{u^2(\tau_1+\tau_2)} &= \sum_{r\in\recur}P(r)\sum_{r'\in\recur, v\in\setpaths(a_1^{\tau_1+\tau_2}r\to r')}P(v)u(v)^2\\
		&=\sum_{r}P(r)\sum_{r',v\in\setpaths(a_1^{\tau_1}r\to r')} P(v) \sum_{r''\in\recur, v'\in\setpaths(a_1^{\tau_2}r'\to r'')}P(v')(u(v)+u(v'))^2\\
		&=\sum_{r,r',v}P(r)P(v)u(v)^2 \sum_{r'',v'}P(v') +\sum_{r,r',v}P(r)P(v) \sum_{r'',v'}P(v')u(v')^2\\
		&+2\sum_{r,r',v}P(r)P(v)u(v)\sum_{r'',v'}P(v')u(v')\\
		&=\avrg{u^2(\tau_1)}+\avrg{u^2(\tau_2)}+2\avrg{u(\tau_1)u(\tau_2)}
	\end{align*}
	The rest of the construction remains the same in virtue of Proposition \ref{prop: abelian is linearity}.
\end{proof}

%% file: boundedvar.tex
\section{Bounded central moment of integrated transport observables}

Here we are interested in the same analysis as the one displayed in \cite{garcia-millan_correlations_2018}.

\begin{proposition}[Bounded central moment of $O(\tau)$]\label{prop: bound central moment O(tau)}
	For any driven-dissipative abelian sandpile, any of the $n$-th central moment of $O(\tau)$ (where $O$ stands for $O$ut), the integrated number of particles dissipated along $\tau$ avalanches, is bounded from above and below.
	In other words, 
	\begin{equation}
		\min(0,(-C)^n)\leq \avrg{\big(O(\tau)-\avrg{O(\tau)}\big)^n}\leq C^n
	\end{equation}
	with $C<+\infty$ the maximal number of particles contained in a stable configuration.
\end{proposition}
\begin{proof}
	The result is a pure consequence of the finiteness of the system, the boundedness of the number of particles inside of it and the control of the particle content through injection and dissipation.
	Without loss of generality we fix $n=1$.
	
	First, notice that $\avrg{O(\tau)}=\tau$ in virtue of the stationary and driving (intermittent dynamic) condition: each particle injected should be dissipated in average.
	
	For the upper bound, we want to overestimate the  maximal fluctuation of $O(\tau)-\avrg{O(\tau)}$.
	Call by $C$ the maximal number of particles that can be accumulated in the system and denote by $r_{\max}$ the associated maximal configuration.
	Conversely, denote by $r_{\min}$ the minimal stable configuration, which we suppose to contain $0$ particles (recalling that we set $z(v)\geq 0$ for any site $v\in V$).
	Remark that after $\tau$ avalanches, at most $C+\tau$ particles injected can be dissipated, supposing that we started initially in $r_{\max}$.
	One therefore gets
	\begin{equation*}
	\avrg{O(\tau)-\avrg{O(\tau)}}\leq\max(O(\tau))-\avrg{O(\tau)}\leq C
	\end{equation*}
	For the lower bound one has
	\begin{equation*}
		-C=(\tau-C)-\tau\leq\min(O(\tau))-\avrg{O(\tau)}\leq \avrg{O(\tau)-\avrg{O(\tau)}}
	\end{equation*}
	The estimate supposes that we start from $r_{\min}$ and reach $r_{\max}$ in $\tau$ avalanches among which $C$ particles must be stabilised.
	Both upper and lower bounds do not scale with $\tau$, the number of avalanches or particle injected, and depend only on the model maximal number of particles $C$.
	The generalisation to any $n\geq 1$ is easily done by taking the bounds to the power of $n$ .
	When $n$ is even, the lower bound $0$ is trivially obtained by definition of a real number raised to an even power.
\end{proof}
The reason behind this result lies in the definitions.
We control every particle injected inside the system, and so we must have a flow compatible with the size of the reservoir of particles, i.e. the system, which saturates by construction.
In a sens, this integrated number of particles dissipated is a fundamental, and rather tautological, feature shared by all out of equilibrium system in their stationary state.
In our case, it is the number of particles, but it could either be the energy, or any other extensive quantity as long as they are bounded.
In fact, from physical reasons, one could argue that above certain scales of the considered extensive quantity (let's say energy), new mechanisms of evolution and dissipation must emerge and so, for the initial description to be valid, the assumption of boundedness seems fairly reasonable. 
That being said, we want to give a simple condition for the result to hold for a space dependent transport observable.

\begin{lemma}[Bounded central moment of $s(\tau)$] \label{lem: bound central moment s(tau)}
	Take any driven-dissipative abelian sandpile and make the following assumption on the transport observable $s(\tau)$, the integrated avalanche sizes in number of topplings
	\begin{itemize}
		\item[(*)] \emph{irreversibility}: each toppling is associated to at least one particle getting closer by one step (i.e. one edge of the graph $G(E,V)$) to the dissipative boundary;
		\item[(**)] \emph{equidistance}: the site $v_1$ is the farthest from the dissipative boundary and is at equidistance $L-1$ from all of the boundary sites. 
		The distance here is the minimal number of edges between two sites.
	\end{itemize}
	Then any of the $n$-th central moment of $s(\tau)$ is bounded in the stationary state, meaning
	\begin{equation}
	\min(0,\big(-LC\big)^n)\leq \avrg{\big( s(\tau)-\avrg{s(\tau)}\big)^n}\leq \big(LC\big)^n
	\end{equation}
	 where $C$ counts the maximal number of particles in a stable configuration of the model.
\end{lemma}

\begin{proof}
	The hypothesis enforce a proportionality relation between $s(\tau)$ and $O(\tau)$.
	The argument follows the same spirit as in Proposition \ref{prop: bound central moment O(tau)}.
	We focus again on the case $n=1$, the other cases $n\geq 1$ follow easily.
	
	First, because of $(*)$, $(**)$ and that on average one particle is dissipated for each new avalanche triggered on the stationary state, we have necessarily $\avrg{s}=L$.
	Indeed, suppose $\avrg{s}< L$, then the average density of the initial state must increase which contradicts the stationary condition. 
	On the other hand, supposing $\avrg{s}>L$, we must then decrease the density of the initial state which is also impossible from stationarity. 
	
	Now, performing $\tau$ avalanches, with an associated integrated current $s(\tau)$, we show that 
	\begin{equation*}
		\avrg{s(\tau)-\avrg{s(\tau)}} \leq \max(s(\tau))-\avrg{s(\tau)} \leq L(\tau+C)-L\tau = LC
	\end{equation*}
	We want an overestimate on the largest value of $s(\tau)$. 
	Naively, it suffices to overestimate the transition from $r_{\max}$ to $r_{\min}$ between the maximal and minimal configurations. 
	Since we imposed $z(v)\geq 0$ for all vertices $v$ of the models that we consider, there is at most $C$ particles in $r_{\max}$ that can be dissipated and at least $0$ particles in $r_{\min}$. 
	Suppose all $\tau+C$ particles are stacked at $v_1$, and imagine you perform $L(\tau+C)$ topplings.
	Then, from $(*)$, it is necessary that each of the $\tau+C$ particles is closer to the dissipative boundary by $L$ jumps.
	From $(**)$, it is clear that each particle has reached the dissipative boundary and we are left with a configuration having no particles.
	Therefore, any avalanche from $a_1^{\tau}r_{\max}$ must involve at most $L(\tau+C)$ topplings.
	
	For the lower bound we can proceed in the same way with the opposite transition from $a_1^{\tau}r_{\min}$ to a virtual maximal configuration where $C$ particles are set on site $v_1$. 
	Along the procedure, we dissipate only $\tau-C$ particles and have an associated toppling cost $(\tau-C)L$. 
	This gives 
	\begin{equation*}
		-LC\leq \min(s(\tau))-\avrg{s(\tau)}\leq \avrg{s(\tau)-\avrg{s(\tau)}} 
	\end{equation*}
\end{proof}
We gave sufficient conditions ensuring that the $n$-th central moments of $s(\tau)$ are bounded.
Replacing $s(\tau)$ by any observable $u(\tau)$ for which $(*)$  and $(**)$ can be adapted must certainly have bounded central moments. 
However, it would be desirable to give a more general statement on the necessary and sufficient conditions on $u(\tau)$ after which one can be sure of the boundedness of $\avrg{(u(\tau)-\avrg{u(\tau})^n}$.
Also, remark that Lemma \ref{lem: bound central moment s(tau)} does not ensure that $\avrg{s^2(\tau)}-\avrg{s(\tau)}^2$ tends to a constant, which is a stronger statement.
Yet it will be sufficient to derive a certain number of non-trivial results.

In \cite{garcia-millan_correlations_2018}, the authors treated the case of $s(\tau)$ in the Oslo model in $1$ and $2$ dimensions.
Here, we can at least apply Lemma \ref{lem: bound central moment s(tau)} to the DSS and the Oslo model in any dimension as long as the condition $(**)$ can be satisfied. 
Also, we find that the Manna model of Definition \ref{def: Manna} also falls inside of the result.
Indeed
\begin{proposition}
	The $n$-th central moment of the integrated avalanche size $s(\tau)$ (number of topplings) in the Manna model of Definition \ref{def: Manna} is bounded for all $\tau$.
\end{proposition}
\begin{proof}
	To prove this statement, we need to verify that the hypothesis of Lemma \ref{lem: bound central moment s(tau)} are verified.
	From the Definition \ref{def: Manna}, one can see that the distance in number of topplings from the injection to any dissipative site is a constant $L$, so $(**)$ is verified.
	More, each toppling occurring on the sites intersecting the axis in the natural direction of $\Z[d]$ and passing by $v_1$ is associated with $2$ particles getting closer to a dissipative site when the particles are transferred along the perpendicular directions of the axis considered.
	At the same positions, when the transfer occurs along the axis and not perpendicularly, only one particle gets closer to a dissipative boundary. 
	When a toppling occurs at $v_1$, two particles are getting closer to the boundary in any direction of transfer.
	On all the other sites, one toppling put one particle closer and another one away from the boundary in any direction of transfer.
	In other words, a toppling is an irreversible process pushing at least one particle toward the exit boundaries.
	Consequently, $(*)$ is also satisfied, which concludes the proof.
\end{proof}

%% file: application.tex
\section{Applications} 

We start with a rather general result which applies when the following property is satisfied: for any $r\in\recur$ we have
\begin{equation}\label{eq: markov irreducible aperiodic}
	W^t\ket[r]\xlongrightarrow[t\to+\infty]{}\ket[\psi]
\end{equation}
In the dedicated literature on Markov chains \cite{levin_markov_2007} we found that if $W$ describes an irreducible and aperiodic Markov process, then it implies Equation \eqref{eq: markov irreducible aperiodic}.
Irreducibility means that it is possible to find a $t\geq 1$ such that $P(a_1^t r \to r')>0$ for any $r,r'\in\recur$.
All recurrent configurations are then connected and can be reached from one another in a finite number of avalanches.
The aperiodicity condition is a bit more abstract and applies when $\textrm{gcd}\{t:~P^t(r,r)>0\}=1$ for all $r\in\recur$, where $\textrm{gcd}$ stands for the greatest common divisor.
Proving, or disproving, these two conditions for each sandpile model is a task on its own that we do not wish to develop here.

\begin{proposition}[Asymptotic time decorrelation of transport observables]\label{prop: asymptotic time decorrelation}
	Take any abelian sandpile for which Equation \eqref{eq: markov irreducible aperiodic} is satisfied for any $r\in\recur$.
	Then, starting from the stationary state $\ket[\psi]$, one has 
	\begin{equation}
		\avrg{uW^tu}\xrightarrow[t\to\infty]{}\avrg{u}^2
	\end{equation}
	for any transport observable $u$.
\end{proposition}
\begin{proof}
	From the hypothesis, we see that
	\begin{align*}
	\bra[\Id]uW^t\ket[r] \xrightarrow[t\to\infty]{} \bra[\Id]u\ket[\psi]=\avrg{u}
	\end{align*}	 
	and 
	\begin{align*}
	\bra[\Id]uW^tu\ket[\psi]=\sum_{r\in\recur}\bra[\Id]uW^t\ket[r]\bra[r]u\ket[\psi] \xrightarrow[t\to\infty]{} \sum_{r\in\recur}\bra[\Id]u\ket[\psi]\bra[r]u\ket[\psi]=\bra[\Id]u\ket[\psi]\sum_{r\in\recur}\bra[r]u\ket[\psi]=\avrg{u}^2
	\end{align*}	
\end{proof}

\begin{lemma}[Time transport correlation]\label{lem: ransport correlations in time}
	Take any abelian sandpile for which  Equation \eqref{eq: markov irreducible aperiodic} is satisfied for any $r\in\recur$.
	Suppose the variance of the integrated transport observable $u(\tau)$ to be bounded.
	Then we must have 
	\begin{equation}\label{eq: variance and time correlation relation}
		\frac{\avrg{u^2}-\avrg{u}^2}{2}=-\sum_{t=0}^{\infty}(\avrg{uW^tu}-\avrg{u}^2)
	\end{equation}
	As a consequence, time correlations of $u$ are mostly negative in the sens that
	\begin{equation}\label{eq: inequality integrated time correlation}
		\sum_{t=0}^{\infty}(\avrg{uW^tu}-\avrg{u}^2)<0
	\end{equation}
\end{lemma}

\begin{proof}
	From Proposition \ref{prop: asymptotic time decorrelation} we see that the integration of the time correlations for non equal times gives
	\begin{equation*}
		\sum_{t=0}^{+\infty}(\avrg{uW^tu}-\avrg{u}^2) = l
	\end{equation*}	
	where $l$ is either finite or infinite. 
	From Lemma \ref{lem: recursion transport corr of ASM}, setting $(\tau_1,\tau_2)=(\tau,1)$, we have the one step recursion 
	\begin{align*}
		\avrg{u^2(\tau+1)}-\avrg{u(\tau+1)}^2 
	=\avrg{u^2(\tau)}-\avrg{u(\tau)}^2 +\big[\avrg{u^2}-\avrg{u}^2+ 2\sum_{t=0}^{\tau}(\avrg{uW^tu}-\avrg{u}^2)\big]
	\end{align*}
	for all $\tau$. 
	However, because we assumed $\avrg{u^2(t)}-\avrg{u(t)}^2 $ to be bounded for all $t\geq 1$, $l$ must be finite and the term in the brackets in the previous expression equals to $0$.
	This gives the condition \eqref{eq: variance and time correlation relation}.
	This also implies \eqref{eq: inequality integrated time correlation} which is given from the positivity of the l.h.s. of \eqref{eq: variance and time correlation relation}. 
\end{proof}
We believe that this result encompasses a great number of known sandpile models.
Yet, even if true, in general the only observable on which it can be applied safely is  $O(\tau)$, the number of particles dissipated from $\tau$ successive avalanches.
One can then wonder if the above result can apply to any transport observable in any settings of abelian sandpile models \emph{without} explicitly supposing an upper bound on the variance of $u(\tau)$.

From the work of \cite{garcia-millan_correlations_2018}, it already appeared that a uniform rate of injection was able to disprove this statement in the case of the Oslo model.
Indeed, for the latter model, it was shown that the system variance of $s(\tau)$ grows linearly with $\tau$ with a uniform drive in space, whereas it saturates to a constant when the injection occurs at the first site of the linear chain in $1$D, or the corner of a finite square lattice in $2$D.
In other words, the drive is a relevant perturbation for this property.
A more detailed discussion on these questions is given in \cite{garcia-millan_correlations_2018} and we invite the interested readers to have a look at it.
Nonetheless, it could be interesting to develop rigorous estimates on the scaling of the variance of some $u(\tau)$, especially investigating its dependence on the shape of the driving.
The DSS might be a good candidate, considering its renewed tractability (from the present paper and \cite{lallemant_space-time_2025}). 
We must stress that taking in a systematic way the realisations of the driving along the computations seems not to be an easy task.

In addition, we notice that as long as $\avrg{u^2(\tau)}-\avrg{u(\tau)}^2$ grows as or more slowly than $\tau(\avrg{u^2}-\avrg{u}^2)$, 
we must have
\begin{equation}\label{eq: general inequality integrated time correlation}
	\sum_{t=0}^{\infty}(\avrg{uW^tu}-\avrg{u}^2)\leq 0
\end{equation} 
which would extend the result of Equation \eqref{eq: inequality integrated time correlation}.
The linear growth $\tau(\avrg{u^2}-\avrg{u}^2)$ emerges, for example, in systems where the RV $u(\tau)=\sum_{t=1}^{\tau}u_t$ is the sum of IID RV $(u_t)_t$.
Finding the hypothesis under which Equation \eqref{eq: general inequality integrated time correlation} is satisfied in sandpile models seems to be an important question.
We conjecture that as long as $u(\tau)$ is positively correlated with $O(\tau)$ (i.e. positive covariance), Equation \eqref{eq: general inequality integrated time correlation} should be true.

\begin{lemma}\label{lem: exact solution transport time correlations}
	Consider any abelian sandpile of a finite size in any dimension set to be initially in their stationary state. 
	Take $u(\tau)$ and suppose its $n$-th central moment to be bounded.
	Suppose also that there exists a finite time $\tau^*\leq t$ such that $\avrg{uW^tu}=\avrg{u}^2$ where $\tau^*$ is called the full mixing time.
	Then, the time correlator $\avrg{uW^tu}$ satisfies the linear system of equations $Y = MX$ where 
	\begin{align}
		Y = \begin{pmatrix}
		g(1)\\
		g(2)\\
		g(3)\\
		...\\
		g(\tau^*)
		\end{pmatrix},
		~M = \begin{pmatrix}
		1& 1& 1& ...& 1\\
		1& 2& 2& ...& 2\\
		1& 2& 3& ... & 3\\
		1& 2& 3& ... & 4\\
		...& ...& ...& ...& ...\\
		1& 2& 3& ... & \tau^*
		\end{pmatrix},
		~X = \begin{pmatrix}
		\avrg{uu}\\
		\avrg{uWu}\\
		\avrg{uW^2u}\\
		...\\
		\avrg{uW^{\tau^*-1}u}
		\end{pmatrix}
	\end{align}
	with 
	\begin{equation}\label{eq: g expression}
		g(t) = t(\tau^*+\frac{1}{2})\mu^2-\frac{\avrg{u^2(t)}}{2}
	\end{equation}
	The inverse $M^{-1}$, defined such that $M^{-1}M=\Id$, is the (almost) tridiagonal matrix with off-diagonals $-1$ and diagonal $2$ with an exception for the last index $\tau^*$ for which $[M]_{\tau^*,\tau^*}=1$ .
	The solution is simply given by
	\begin{equation}\label{eq: solution time correlations}
		\avrg{sW^ts}=\left\{\begin{array}{ll}
		 2g(1)-g(2) &\text{for}~t=0\\
		 2g(t+1)-g(t+2)-g(t) &\text{for}~1\leq t\leq \tau^*-2\\
		 g(\tau^*+1)-g(\tau^*) &\text{for}~t=\tau^*-1
		\end{array}\right.
	\end{equation}
\end{lemma}
\begin{proof}
	Notice that we are in the same situation as in Lemma \ref{lem: ransport correlations in time} with the additional constraint 
	\begin{equation*}
		\avrg{uW^ku}-\avrg{u}^2 = 0
	\end{equation*} 
	for all $k\geq \tau^*$.
	From Equation \eqref{eq: recursion transport corr of ASM}, the condition of finite full mixing time $\tau^*$ fixes, for $(\tau_1,\tau_2)=(\tau^*,t)$, the relation
	\begin{equation*}
	\avrg{u^2(t)}-\avrg{u(t)}^2+ 2\big[\sum_{k=0}^{t-1}(k+1)\avrg{uW^{k}u}
	+ t\sum_{k=t}^{\tau^*-1}\avrg{uW^{k}u}-\big(\tau^*t-\frac{(t-1)t}{2}\big)\mu^2\big]=0
	\end{equation*}
	which translates in the linear system
	\begin{equation*}
		t(\tau^*+\frac{1}{2})\mu^2-\frac{\avrg{u^2(t)}}{2} = \sum_{k=0}^{t-1}(k+1)\avrg{uW^{k}u}+ t\sum_{k=t}^{\tau^*-1}\avrg{uW^{k}u}
	\end{equation*} 
	It is then an easy task to check the condition $M^{-1}M=\Id$ computing the first entries of the matrix product.
\end{proof}

The above lemma shows that, under the suitable assumptions, time transport correlations and integrated transport over time form a closed system of equations. 
In fact, we can think of at least $3$ different $1$D abelian sandpile models for which the formula applies.
Namely, it applies at least to $O(\tau)$, $s(\tau)$ and $s_v(\tau)$ (local number of topplings at site $v\in V$) observables for the DSS and the Oslo model; and at least to $O(\tau)$ for the ARW of Definition \ref{def: ARW}.
If the assumption of a finite full mixing time was already proven to hold for the first two models \cite{dhar_steady_2004, lallemant_space-time_2025}, for the ARW we must give hereafter a proper argument.
\begin{proposition}
	The ARW of Definition \ref{def: ARW} satisfies $W^L\ket[r]=\ket[\psi]$ for any $r\in\recur$.
\end{proposition}
\begin{proof}
	In the ARW of Definition \ref{def: ARW}, the driving is at site at $x=1$, and we have a left reflexive and right dissipative boundaries.
	It now suffices to check that their exists $k$ such that the minimal configuration $\ket[0^L]$ satisfies $W^k\ket[0^L]=\ket[\psi]$. 
	From the model rules, it is then a formality to extend the result starting from any initial recurrent configuration $r\in\recur$.
	Let's start by adding $k=2$ particles. 
	Using the abelianess, we can see that 
	\begin{align*}
		a_1^2\ket[0^L] &\to pa_1\ket[10^{L-1}]+qa_1a_2\ket[0^L]= pa_1^2\ket[10^{L-1}]+qa_1a_2\ket[0^L]\\
		\Leftrightarrow a_1^2\ket[0^L] &\to q\sum_{t=0}^{\infty}p^t a_1a_2\ket[0^L]=\frac{q}{1-p}a_1a_2\ket[0^L]=a_1a_2\ket[0^L]
	\end{align*}
	Similarly, we show that $a_1^3\ket[0^L]\to a_1^2a_2\ket[0^L]\to a_1a_2a_3\ket[0^L]$.
	By induction, this proves
	\begin{equation}
		\evol^{\infty}(a_1^L\ket[0^L]) = \evol^{\infty}(\prod_{x=1}^{L}a_x\ket[0^L])=\evol^{\infty}(a_x^k\ket[0^L])
	\end{equation}
	for any $k\geq L$. 
	Therefore, $\evol^{\infty}(a_1^L\ket[0^L])$ is none but the stationary state $\ket[\psi]$ of the model. 
\end{proof}
Notice that when the ARW is set in its original driven-dissipative setup, meaning with both boundaries being dissipative, the full mixing time is infinite and so Lemma \ref{lem: exact solution transport time correlations} does not apply. 

\begin{remark}
	Lemma \ref{lem: exact solution transport time correlations} provides a definitive answer to one of the main problem left in \cite{lallemant_space-time_2025}, which was precisely to find a simple expression for the the time correlations of the avalanche size $\avrg{sW^ts}$.
	We have just demonstrated that $\avrg{sW^ts}$, in the DSS of Definition \ref{def: DSS}, equals to the Laplacian (with respect to $t$) of a function $g(t)$ which is linear in the second moment of the integrated avalanches size $\avrg{s^2(t)}$.
	From \cite{lallemant_space-time_2025}, the quantity $\avrg{s^2(t)}$ can be simply interpreted: it is none but the average of the square of the area enclosed by a $1$D symmetric random walker in a specific setup.
	The walker propagates to the right by one unity at each time step and can perform three movements in the vertical direction: $\{-1,0,1\}$. 
	The walker starts at time $0$ at the position (height) $t$ and experiences two absorbing walls: one is horizontal and is set at the height $0$, the other is vertical and set at time $L+1$. 
	Also the time correlator $\avrg{sW^ts}$ takes the same values when exchanging $p\leftrightarrow q$, as the random walk is symmetric \cite{pruessner_exact_2004} with respect to this transformation of the microscopic probabilities.
	We verified the solution up to $L=9$ by comparing straightforward matrix multiplication $\bra[\Id]sW^ts \ket[\psi]$ with the result obtained using the relation \eqref{eq: solution time correlations} and computing $\avrg{s^2(t)}$.
\end{remark}

Finally, we must stress that the above construction can be extended to get a closed expression for the $n$-points time correlator function. 
However, it gets quickly complicated to deal with the corresponding system of equations in a systematic way, and we do not intend to derive an explicit form in this paper.
From our current understanding of the problem and first attempts in this direction, we believe that it should be a work on its own.
Notice that Lemma \ref{lem: exact solution transport time correlations}, and the natural extension to the $n$-points correlation in time, is reminiscent of the theory of integrability in equilibrium system and the existence of underlying symmetries and conservation laws.
In fact, it is clear that the central moments of $O(\tau)$ (Proposition \ref{prop: bound central moment O(tau)}) are the fundamental conserved quantities behind the integrability of the transport observables that we identified to be "proportional" to $O(\tau)$ (e.g. Lemma \ref{lem: bound central moment s(tau)}).
We must emphasize that these conserved quantities are of a \emph{statistical nature} and that conservation holds only, and strictly, for time $t\geq \tau^*$, where $\tau^*<+\infty$ is model dependent.
The other important condition of Lemma \ref{lem: exact solution transport time correlations}, that allows to close the set of equations \eqref{eq: recursion transport corr of ASM}, is the complete loss of memory in the system in finite time (finite full mixing time).
It could be enlightening to further dig into the relations between the previous statements and the known approach to out of equilibrium integrable systems (e.g. Generalised Gibbs Ensemble).

%% file: conclusion.tex
\section{Conclusion}

In this paper, we provide an extension of known results and derive a number of new ones on time transport correlations in abelian sandpile models.
The argumentation uses simple tools and concepts.
The number of hypothesis are also only a few, explaining the range of models captured by the approach.

We first use the abelianity of the models to decompose the response of a super avalanche into a sum of smaller avalanches (Equation \eqref{eq: lineartiy of <u(tau)>}). 
This simple construction allows then to derive a recursion on the second moment of integrated transport in terms of simple contributions (Equation \eqref{eq: recursion transport corr of ASM}).
From there, inspired by the work of \cite{garcia-millan_correlations_2018}, we give sufficient conditions upon which the $n$-th central moment of the integrated size of $\tau$ successive avalanches $s(\tau)$ is bounded for all times. 
This boundedness is then coupled to the recursion (Equation \eqref{eq: recursion transport corr of ASM}) to provide a simple relation in a given class of abelian sandpile models.
The relation in question connects the integral of the $2$-points time correlation of transport observables to their variance (Equation \eqref{eq: variance and time correlation relation}).
This point out toward a duality relation between time dependent and equal time properties.
Different models are discussed and related to this result, establishing its general applicability. 
Finally, by adding an additional hypothesis on the models, we manage to close the recursion and are just left with a linear system of equations.
We solve this system and find that the transport $2$-points time correlator equals to the Laplacian (in time) of a function that depends linearly on the second moment of the integrated transport over time.

On a more model dependent basis, one of the main achievement, from the partial point of view of the author, has been to apply the previously discussed solution to the DSS model.
Indeed, it answers one of the main open questions left in the recent work \cite{lallemant_space-time_2025}. 
It was originally not intended to tackle the problem in this way, and find the final answer to be remarkably simple.
The present work potentially provides important missing pieces of the complete analytical study of the DSS model in its driven dissipative setup, of which integrability seems now pretty clear. 
We expect that many more results and generalisations can be built upon the material of this text, of which some have already been discussed along the argumentation.
We believe also that the approach developed hereby should extend naturally to the (local and global) density correlations in some classes of abelian sandpile models.

\section*{Aknowledgment}
The author wishes to thank Arthur Dedieu and Marilou Dazy for interesting discussions.